# Gravity-induced phase-shift of light : outline of an interferometric test of the Equivalence Principle

Eduardo Díaz-Miguel

Departmento de Física Aplicada I, Facultad de Ciencias
Universidad de Málaga, 29071-Málaga, Spain.

I analyze the change of the interference pattern in an optical interferometer when it passes from rest to free fall. It is shown that the "disconnection" of the gravitational field causes a jump in the phase difference that could be measured with the current sensitivity of these instruments. For this reason, I propose to the optical interferometry community the possibility of a test of the Equivalence Principle based on the aforementioned effect.

## 1. INTRODUCTION

It is well known that many gravitational red-shift experiments have been made since the first successful measurement of Pound-Rebka-Sneider (1960-65). The first one was the Vessot-Levine Rocket Red-Shift Experiment (1978). This tests are founded on the ultrahigh stability of masers cloks, superconducting-cavity stabilized oscillators, etc ( see [1] for an authorized revision).
   Our proposal, [2], [3], is also a test of the Equivalence Principle (EP). However it is based on a phase shift measure. We use the fact that the phase difference on a falling interferometer (Eintein´s elevator) is not equal to the one observed when it is at rest.
   This proposal is a classical counterpart of the well known gravito-inertial induced quantum interference experiments [4] .

## 2. CALCULATION OF THE JUMP OF THE PHASE DIFFERENCE

Figure (1) is the classic scheme of a Michelson interferometer. The laser source F emits an almost monochromatic wave with frequency $\nu$ and wavelength $\lambda$. This wave reaches a half-silvered mirror, E, and splits into two beams which, after reflection in the mirrors E1 and E2, interfere at the point O. Let us denote by $(\Delta\Phi)_{rest}$ the phase difference between the two interfering beams after traveling the lengths $L_1$ and $L_2$, respectively.. Suppose that all the experimental device ( source, mirrors, detectors, etc) passes to a state of free fall. Let $(\Delta\Phi)_{free\ fall}$ bee the phase difference that would be observed in this new situation. Based on the EP, I will prove that the jump of the phase difference between the two interferometric states is given by

$$(\Delta\Phi)_{rest} - (\Delta\Phi)_{free\ fall} = \frac{2\pi}{\lambda} \frac{\Delta L}{c^2 R} \frac{GM}{} \qquad (1)$$

where $\Delta L = L_2 - L_1$, M is the mass of Earth and R its radius. If the time of the fall happens at a distance r, then this is the value that should be replaced, in the above equation, instead of R.

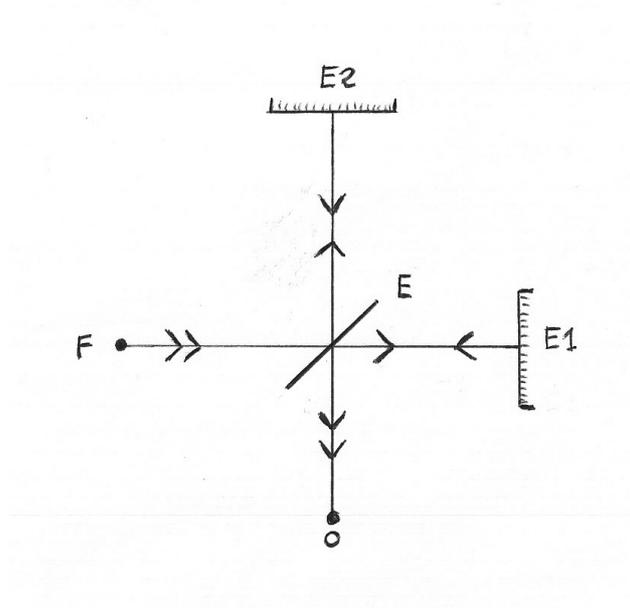

Figure 1. Diagram of a Michelson interferometer..

## 2.1 ELEMENTAL DERIVATION OF (1)

The basic equation of wave interference is: phase difference = (wave number) x (path difference). Therefore

$$(\Delta \Phi)_{rest} = k\Delta L = \frac{2\pi}{\lambda}\Delta L = \frac{2\pi \nu}{c}\Delta L \qquad (2)$$

The frequency appearing in the above equation is the one of the photons propagating on the Earth´s surface. Let us find what would be the phase difference if the photons were not subjected to the gravitational field. To do this we take into account Einstein equation for the red-shift of light:

$$\nu = \nu_0 (1 + \frac{GM}{c^2 R}) \qquad (3)$$

where $\nu_0$ is the frequency of photons at infinity. So

$$(\Delta \Phi)_{rest} = \frac{2\pi \nu_0}{c}\Delta L (1 + \frac{GM}{c^2 R}) \qquad (4)$$

Therefore, the phase difference that would be measured in the absence of gravity is.

$$(\Delta \Phi)_{no\ gravity} = \frac{2\pi \nu_0}{c}\Delta L$$

According to the EP, it is the same as in free fall. So

$$(\Delta \Phi)_{free\ fall} = \frac{2\pi \nu_0}{c}\Delta L \qquad (5)$$

Subtracting (5) from (4) leads to

$$(\Delta \Phi)_{rest} - (\Delta \Phi)_{free\ fall} = \frac{2\pi\ v_0 \Delta L}{c} \frac{GM}{c^2 R} \qquad (6)$$

One might object that in the above equation appears an unmeasurable magnitude: the frequency, $v_0$, of a photon at infinity. However, since the right hand side of (6) is of first order in the parameter $GM/c^2R$, we can replace, because of equation (3), $v_0$ by $v$. This gives the desired result (equation (1)), which only involves controllable variables in the laboratory. The magnitude $GM/c^2r$ is the dimensionless potential. Its value at the Earth´s surface is $6.95 \cdot 10^{-10}$, what justifies the first order approximation made.

## 2.2 A FORMAL PROOF

Let $(k^0, \mathbf{k})$ be the wave vector of a light ray that travels from points P to Q. The change that undergoes the phase $\phi$ of a photon in the geometrical optics limit is [5]

$$\Delta \Phi = -\int_\Gamma k_\alpha\ dx^\alpha\ ,\quad \alpha = 1,2,3 \qquad (7)$$

where $\Gamma$ is the null geodesic that joins P and Q. In a static gravitational field [6]:

$$k_\alpha = \frac{\omega_0}{c} \frac{g_{\alpha\beta}}{\sqrt{g_{00}}} \frac{dx^\beta}{dl} \qquad (8)$$

$$ds^2 = g_{00} c^2 dt^2 - dl^2 \qquad (9)$$

$$dl^2 = - g_{\alpha\beta}\ dx^\alpha\ dx^\beta \qquad (10)$$

where $\omega_0 = -\dfrac{\partial \eta}{\partial t}$.

The eikonal, $\eta$, is independent of t in a static field. So $\omega_0$ is constant along the ray. Let $\omega$ be the (varying) proper frequency. We have

$$\omega = \frac{\omega_0}{\sqrt{g_{00}}} \qquad (11)$$

Therefore $\omega_0$ is the frequency at infinity. Inserting (8) and (10) in (7) and considering (11) and the fact that for a light ray $ds^2 = 0$, we find

$$\Delta \Phi = \int_\Gamma \omega\ d\tau = \omega_0 \int_\Gamma dt = \omega_0 \Delta t \qquad (12)$$

where $\Delta t$ is the interval of time coordinate and $\tau$ is the proper time measured by the observers situated along the path of the light ray.

To find the time travel of light we follow a straight-line approximation, valid to first order in $GM/c^2R$ [6]: the differential equation for a light ray in the first order Schwarzschild metric is (in the equatorial plane):

$$(1 - \frac{2GM}{c^2 r}) c^2 dt^2 - (1 + \frac{2GM}{c^2 r}) dr^2 - r^2 d\varphi^2 = 0$$

We will use the configuration of figure (2) (greatly exaggerated) : the interferometer arm is horizontal and the Schwarschild radial coordinates of the points P and Q are R and $r_0$ respectively. Now we approximate the actual path of the photon by a straight line: $r \cos\varphi = R$. The result of the integration is

$$\Delta t = \frac{1}{c}\left[\sqrt{r_0^2 - R^2} + \frac{2GM}{c^2}\ln\left(\frac{r_0 + \sqrt{r_0^2 - R^2}}{R}\right) - \frac{GM}{c^2}\frac{\sqrt{r_0^2 - R^2}}{R}\right] \quad (13)$$

In our experimental proposal, P and Q are the ends of an interferometer arm whose proper length is L. This is a physical constraint that must be taken into account. To do so, we effectuate another straight-line approximation using the spatial part of the metric:

$$dl^2 = (1 + \frac{2GM}{c^2 r})dr^2 + r^2 d\varphi^2$$

Integrating, we arrive at the following constraint

$$L = \int_P^Q dl = \sqrt{r_0^2 - R^2} + \frac{GM}{c^2}\ln\left(\frac{r_0 + \sqrt{r_0^2 - R^2}}{R}\right) - \frac{GM}{c^2}\frac{\sqrt{r_0^2 - R^2}}{R} \quad (14)$$

If we compare (13) and (14), then

$$\Delta t = \frac{1}{c}\left[L + \frac{GM}{c^2}\ln\left(\frac{r_0 + \sqrt{r_0^2 - R^2}}{R}\right)\right] \quad (15)$$

To first order, we can make in the second term of the right-hand of (15) the (zeroth order approach) substitution

$$\sqrt{r_0^2 - R^2} \approx L$$

For $L/R \ll 1$, equation (15) gives

$$\Delta t = \frac{L}{c}(1 + \frac{GM}{c^2 R})$$

So equation (12) reads

$$\Delta \Phi = \frac{\omega_0 L}{c}(1 + \frac{GM}{c^2 R}) = \frac{2\pi \nu_0}{c} L (1 + \frac{GM}{c^2 R})$$

according to equation (4) of our previous derivation.

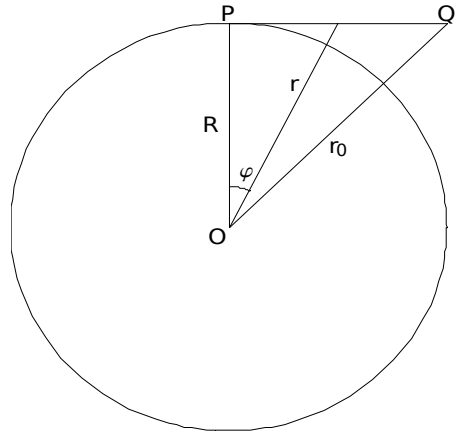

Figure 2. Quantities referred to in the calculation of the phase difference.

## 3. EFFECT OF TIDAL FORCES

We have tacitly assumed that our falling platform is a perfect locally inertial system. Obviously this is not so. It has a finite size and the length of the fall is also finite. There are tidal forces that compress the arms. Let us consider a uniform rod of length l, density $\rho$ and Young modulus Y. When the rod falls freely in the horizontal position and travels the distance H, an easy estimation of the relative contraction gives

$$\frac{\delta l}{l} \approx \frac{3\rho \, g \, l^2 H}{Y \, r^2}$$

where g is the gravity acceleration at a distance r.

For example, when l = H = 1m, the above equation gives, for an Aluminum rod and on the Earth´s surface, a relative contraction of the order of $10^{-17}$. This effect is about a million time less than the one we want measure: recall that
$GM/c^2 R = 6.95 \cdot 10^{-10}$.

## 4. EXPERIMENTAL PROPOSAL

One of the most famous research centers related to free fall experiments is the "Bremen Drop Tower", which belongs to ZARM (Center for Applied Space Technology and Microgravity). This tower provides researchers 4.74 seconds of microgravity, three times a day, for studies of the behavior of various physical systems, mainly hydrodynamics. The aim of ZARM to provide the maximum amount of time controlled weightlessness. However, in my proposal, to get a jump on the phase difference it is not necessary such a long time. This jump is, for experimental purposes, instantaneous: what matters is not the duration of the interferometer free fall, but it occurs; even for a split second. Using an electromagnetic analogy: to induce an electric current in a loop it is not necessary to move a magnet for long. It is sufficient that there is a variation of the magnetic field flux across the loop for an arbitrarily short time.

We can then imagine an interferometer that is being released from his situation of rest to a state of free fall, during which could be a few tenths of a second. As the maximum speed reached is small enough, the braking system need not be so complex as in the ZARM. In this regard, it should be noted that there are built with fiber optic interferometers that have a great mechanical strength due to the possibility of including in the same substrate the emitter, waveguide and detector.

Figure (3) represents what might be the sequence of one of these experiments. The increasing and decreasing curves of the variation of the phase difference, $\delta\Phi$, are indicated by dashed lines, since their shapes depend on the type of control that is exercised over the interferometric platform. Obviously, the maximum value of said change, given by equation (1), is independent of the manner in which this control is done.

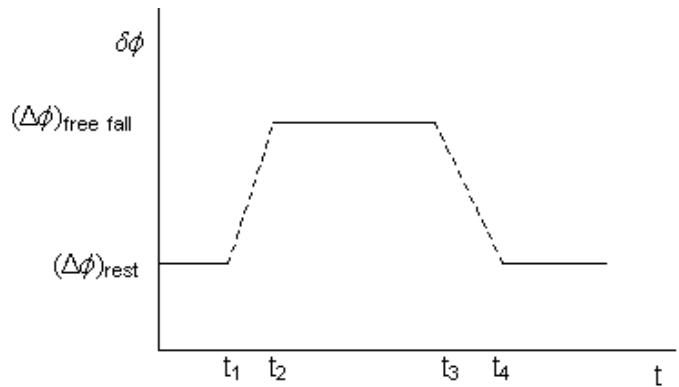

Figure 3. Variation in time of the phase difference, $\delta\Phi$, in a hypothetical free fall interferometric experiment. At time $t_1$ begins the release of the devices that keep it at rest. The free fall occurs in the interval $[t_2, t_3]$. The braking interval is $[t_3, t_4]$. The phase jump, $(\Delta\Phi)_{free\ fall} - (\Delta\Phi)_{rest}$, is given by equation (1).

## 5. CONCLUSION

This article is only a sketch. Not being an expert in experimental techniques, I can not discuss mechanical, thermal or optical instabilities of a falling interferometric platform. These are the issues that decide, ultimately, if the experiment is feasible or not. However, there is no doubt that, if this experiment could be performed, would become another nice confirmation of the EP.

**REFERENCES**

[1] WILL, C. M. *Theory and experiment in Gravitational Physics*. Cambridge University Press (1993).

[2] DÍAZ-MIGUEL, E. *The Principle of Equivalence and the gravity-induced phase-shift: an outline of experimental proposal*. Spanish Relativity Meeting (ERE). Universitat de les Illes Ballears. Spain, (1997).

[3] DÍAZ-MIGUEL, E. *The Principle of Equivalence and the gravity-induced phase-shift: an outline of experimental proposal* . Anales de Física **95** (2000) 225-228.

[4] AUDRETSCH, J., HEL, F.W., LAMMERZHAL, C. *Matter Wave Interferometry and why Quantum Objects are fundamental for Establishing a Gravitational Theory*. Relativistic Gravity Research. Lectures Notes in Physics 410. Springer Verlag (1991).

[5] LANDAU, L.D. , LIFSHITZ, E. M. *The classical theory of fields*. Volume 2. Pergamon Press (1975).

[6] D´INVERNO, R. *Introducing Einstein´s Relativity*. Clarendon Press. Oxford (1990).